\documentclass[anonymous,runningheads]{llncs}
\usepackage{graphicx}
\usepackage{fullpage}
\usepackage{algorithm}
\usepackage[noend]{algpseudocode}
\usepackage{amsmath}
\usepackage{multirow}
\usepackage{graphicx}
\usepackage{xcolor}
\usepackage{amsfonts}
\usepackage{pifont}
\usepackage{xspace}
\usepackage{mathtools}
\usepackage{enumitem}
\usepackage{lineno}
\usepackage{thmtools}
\usepackage{thm-restate}
\usepackage{relsize}
\usepackage{hhline}
\usepackage{etoolbox}
\usepackage{thmtools}
\usepackage{thm-restate}
\usepackage{array}
\usepackage{cleveref}
\usepackage[most]{tcolorbox}
\usepackage{multicol}
\usepackage{changepage}
\usepackage{algpseudocode}
\usepackage{etoolbox}
\usepackage{relsize}
\usepackage{scalefnt}
\nolinenumbers 

\usepackage{wrapfig}

\setlist[description]{leftmargin=*,labelindent=*}

\newcommand{\alglinenoNew}[1]{\newcounter{ALG@line@#1}}
\newcommand{\alglinenoPop}[1]{\setcounter{ALG@line}{\value{ALG@line@#1}}}
\newcommand{\alglinenoPush}[1]{\setcounter{ALG@line@#1}{\value{ALG@line}}}

\newtheorem{observation}{Observation}{}

\newcommand{\mypara}[1]{\smallskip\noindent\textbf{#1.}}

\newcommand{\ia}{\textit{i}}
\newcommand{\ib}{\textit{ii}}
\newcommand{\ic}{\textit{iii}}

\newcommand{\com}[1]{}

\algdef{SE}[Receiving]{Receiving}{EndReceiving}[1]{\textbf{upon
		receiving}\ #1\ \algorithmicdo}{\algorithmicend\ \textbf{}}%
\algtext*{EndReceiving}

\algdef{SE}[Upon]{Upon}{EndUpon}[1]{\textbf{upon}\ #1\ \algorithmicdo}{\algorithmicend\ \textbf{}}%
\algtext*{EndUpon}

\newcommand\StateX{\Statex\hspace{\algorithmicindent}}

\algrenewcommand\textproc{}

\newcommand{\temph}[1]{\textbf{#1}}

\makeatletter \let\sv@thm\@thm \def\@thm{\let\indent\relax\sv@thm} \makeatother

\newcommand{\calF}{\mathcal{X}}

\crefname{table}{table}{tables}
\crefname{table}{Table}{Tables}
\crefname{algocf}{alg.}{algs.}
\crefname{algocf}{Alg.}{Algs.}
\crefname{figure}{Fig.}{Figs.}
\crefname{figure}{fig.}{figs.}
\crefname{claim}{claim}{claims}
\crefname{claim}{Claim}{Claims}
\crefformat{chapter}{\S#2#1#3}
\crefmultiformat{chapter}{\S\S#2#1#3}{and~#2#1#3}{, #2#1#3}{, and~#2#1#3}

\crefformat{section}{\S#2#1#3}
\crefmultiformat{section}{\S\S#2#1#3}{and~#2#1#3}{, #2#1#3}{, and~#2#1#3}

\usepackage{enumitem}
\setlist{nosep} 
\setlist{itemsep=1pt, topsep=3pt}

\newcommand{\sys}{Grassroots Flash\xspace}

\newcommand{\CD}{Cordial Dissemination\xspace}

\begin{document}

\title{\sys:\\ A Payment System for Grassroots Cryptocurrencies}

\author{Andrew Lewis-Pye\inst{1} \and 
Oded Naor\inst{2}\and 
Ehud Shapiro\inst{1,3}
}%
\authorrunning{Lewis-Pye, Naor, Shapiro}
%
\institute{London School of Economics \and
Technion and StarkWare \and
Weizmann Institute of Science}
\maketitle              
%
\begin{abstract}
The goal of \emph{grassroots cryptocurrencies} is to provide a foundation with which local digital economies can emerge independently of each other and of global digital platforms and global cryptocurrencies; can form and grow without initial capital or external credit; can trade with each other; and can gradually merge into a global digital economy.
Grassroots cryptocurrencies turn mutual trust into liquidity and thus could be a powerful means for
`banking the unbanked'.

Grassroots cryptocurrencies have not been provided yet with a payment system, which is the goal of this paper.
Here, we present \emph{Grassroots Flash}, a  payment system for grassroots cryptocurrencies that employs the \emph{blocklace}---a DAG-like counterpart of the blockchain data structure.  We analyze its security (safety, liveness, and privacy) and efficiency, and that it is indeed grassroots.

\end{abstract}

\section{Introduction}

\mypara{Background and Overview}
\emph{Grassroots cryptocurrencies}~\cite{shapiro2022gc} are a means for turning mutual trust into liquidity,  and thus offer a novel route to ``banking the unbanked''\cite{agarwal2017banking,dupas2018banking,bruhn2009economic}.
The goal of grassroots cryptocurrencies is to provide a foundation with which local digital economies can emerge independently of global digital platforms and global cryptocurrencies; can form and grow without initial capital or external credit; can trade with each other; and can gradually merge into a global digital economy.  This is in contrast to standard \emph{global cryptocurrencies} (e.g. Bitcoin, Ethereum), where a community that wishes to use them for their internal purposes (e.g. to form a community bank, an example which is elaborated on below) must cough up some capital, e.g. to pay gas.  Grassroots cryptocurrencies can be compared to standard global cryptocurrencies along multiple dimensions, as shown in the table below. The starred$^*$ entries are contributions of this paper.  

The original paper~\cite{shapiro2022gc} presented the concept, scenarios, and mathematical specification of grassroots cryptocurrencies, but did not present a viable implementation.  This is the purpose of the present paper. 

Grassroots cryptocurrencies as a mathematical concept, as well as the implementation presented here, are instances of \emph{grassroots distributed systems}~\cite{shapiro2023grassrootsBA}.
Informally, a distributed system is grassroots if it may have multiple deployments that are independent of each other and of any global resources, and that can later interoperate once interconnected.  Such systems may allow communities to conduct their social, economic, civic, and political lives in the digital realm solely using their members' networked computing devices  (e.g., smartphones), free of third-party control, surveillance, manipulation, coercion, or rent seeking by global digital platforms.

\begin{minipage}{\textwidth}
\smaller
  \centering
\newcounter{ccc}
\begin{tabular}{ | m{13.5em} | m{14em}| m{12.5em} | } 

    \hline
    \textbf{Cryptocurrencies} & \textbf{Global (Bitcoin, Ethereum)}
 &  \textbf{Grassroots}
\\
     \hline
\hline
     \textbf{\stepcounter{ccc}
\theccc. Architecture}  & Global & Grassroots
\\
     \hline  
      \textbf{\stepcounter{ccc}
\theccc. Cryptocurrencies} & 1 per blockchain & 1 per agent (its Sovereign), Mutually-pegged \\
     \hline
     \textbf{\stepcounter{ccc}
\theccc. Actors} & Miners \& Accounts & Sovereign \& Traders \\
    \hline 
  \textbf{\stepcounter{ccc}
\theccc. Supply controlled by} & Protocol  & Sovereign \\
     \hline 
        \textbf{\stepcounter{ccc}
\theccc. Equivocation Exclusion by/ Finality by/ Trust in} & Majority of mining power/stake & Sovereign \\
     \hline 
      \textbf{\stepcounter{ccc}
\theccc. Finality} & Probabilistic & Definite$^*$ \\
     \hline 
       \textbf{\stepcounter{ccc}
\theccc. Miner/Sovereign\newline  requirements} & High performance computing, cheap energy and/or capital  & Smartphone \& friends$^*$ \\
     \hline  
    
     \textbf{\stepcounter{ccc}
\theccc. Trader requirements} & Networked computer, capital  & Smartphone \& friends$^*$ \\
     \hline
      \textbf{\stepcounter{ccc}
\theccc. Interoperability} & External, difficult  & Integral via mutual pegging \\
     \hline 
      \textbf{\stepcounter{ccc}
\theccc. Liquidity by} & Mining/External capital  & Trust-based credit lines across cryptocurrencies  \\
     \hline  
    \textbf{\stepcounter{ccc}
\theccc. Data Structure}  & Blockchain (totally ordered) & Blocklace (partially ordered)
\\
     \hline
      \textbf{\stepcounter{ccc}
\theccc. Global State} & Replicated\footnote{If all agents must obtain and maintain a complete copy of the entire data structure in order to operate it should be called \emph{replicated};  it should be called \emph{distributed} if different agents access, maintain and store different parts of the data structure.  Similarly for  \emph{distributed ledgers} vs. \emph{replicated ledgers}, with blockchains falling under the second category.  Note that sharded blockchains are replicated, as an agent must obtain all shards in order to operate on the blockchain.} & Distributed \\
     \hline
        \textbf{\stepcounter{ccc}
\theccc. Block creation} & Competitive & Decentralized\footnote{By \emph{decentralized} we mean decentralized control,  so that each agent controls only part of the data structure. Contrast this with global cryptocurrencies, where miners vie competitively for control of a single entity: The growing tip of the blockchain.}, Cooperative  \\
     \hline
        \textbf{\stepcounter{ccc}
\theccc. Block dissemination} & Global/All-to-All & Local/Grassroots$^*$ \\
     \hline
        \textbf{\stepcounter{ccc}
\theccc. Consensus-Based} & Yes & No$^*$ \\
     \hline
        \textbf{\stepcounter{ccc}
\theccc. Payment Model} & UTXO, Accounts  & UTXO$^*$ \\
     \hline
       \textbf{\stepcounter{ccc}
\theccc. Business Model} & ICO, Miner remuneration & Public good \\
     \hline
    
      \textbf{\stepcounter{ccc}
\theccc. Implementations} & Plenty & Not yet \\
     \hline
  \end{tabular}
\end{minipage}

In a system of grassroots cryptocurrencies, each currency consist of units of debt, henceforth \emph{coins}, that can be issued by anyone---people, corporations, banks, municipalities and governments, henceforth \emph{agents}---and traded by anyone. Unlike mainstream cryptocurrencies, and similarly to fiat currencies,  each agent has sovereignty over minting domestic coins and may price their goods and services in their domestic currency. Unlike fiat currencies, grassroots cryptocurrencies are mutually-pegged by the principle of \emph{coin redemption}: An agent must agree to redeem any domestic coin it has issued against any foreign coin it holds.  Coin redemption has many ramifications, including fungibility of domestic coins; liquidity via mutual coin exchange;  coin arbitrage; and the 1:1-pegging of mutually-liquid cryptocurrencies.

In this paper, we present \emph{\sys}, a payment system for grassroots cryptocurrencies. 
A \emph{payment system}~\cite{guerraoui2019consensus}, sometimes also known as an \emph{asset transfer system}, is an abstraction that aims to capture  Bitcoin's core functionality. While almost all cryptocurrencies implement some form of consensus to totally order the transactions in the blockchain, Guerraoui et al.~\cite{guerraoui2019consensus} prove that realizing a payment system is a weaker problem:  Payments with no dependencies between them need not be ordered, a payment system does not require consensus and can be realized deterministically in an asynchronous setting. 
Concrete implementations for (non-grassroots)  payment systems have been introduced by a number of papers~\cite{auvolat2020money,collins2020online,lewispye2023flash}.
In this paper, we use ideas presented in \emph{All-to-All Flash}~\cite{lewispye2023flash}, a UTXO-based asynchronous payment system, and adapt them to create a grassroots implementation of payment system for grassroots cryptocurrencies.

We establish safety and liveness for the protocol, analyze its efficiency,  and prove it to be \emph{grassroots} ~\cite{shapiro2023grassrootsBA}.  Informally, a distributed system is grassroots if it is permissionless and can have autonomous, independently-deployed instances---geographically and over time---that may interoperate voluntarily once interconnected. 
The quintessential example of a grassroots system is a serverless smartphone-based  social network supporting multiple independently-budding communities that can merge when a member of one community becomes also a member of another~\cite{shapiro2023gsn}. 
Grassroots cryptocurrencies are not that different:  They should support multiple independently-budding digital economies that can merge when a member of one economy becomes also a member of another.

In~\cite{shapiro2023grassroots}, it is argued that systems that require a server to operate (e.g., cloud-based/client-server) or are designed to have a single global instance, are not grassroots.
Neither are peer-to-peer systems that require all-to-all dissemination, including mainstream cryptocurrencies and standard consensus protocols~\cite{castro1999practical,yin2019hotstuff,keidar2021need}, 
since a community placed in a larger context cannot ignore members of the larger context.
Neither are systems that use a global shared data-structure such as pub/sub systems~\cite{chockler2007constructing}, IPFS~\cite{benet2014ipfs}, and distributed hash tables~\cite{stoica2003chord},  since a community placed in a larger context cannot ignore updates to the shared resource by others. Even the `fediverse'~\cite{fediverse}---the universe of applications based on federated servers that use primarily the ActivityPub protocol~\cite{activitypub}---is not grassroots, since two `client' agents cannot communicate without relying on a third party, a server.  While this may sound like a technicality, it is not, as in practice it implies that agents must compromise their digital sovereignty and privacy: They must share personal information with third parties (server operators) in order to communicate, and even if the actual content of the communication could be encrypted, their social graph as well as metadata on their communications would be visible to, and possibly manipulable by, these third party.
In this paper, we prove that indeed \sys is grassroots.

\vspace{0.2cm} 
\mypara{Roadmap} The rest of the paper is organized as follows: Section \ref{section:gc} provides an overview of grassroots cryptocurrencies.  Section \ref{section:blocklace} provides an overview of the blocklace data structure. Section \ref{section:spec} provides a specification of Grassroots Flash via desiderata, and proves safety and liveness of the protocol, assuming all-to-all dissemination.  Section \ref{section:cordial-dissemination} augments \sys with a Cordial Dissemination  protocol, proves a liveness theorem qualified by the structure of the social graph, and proves that the resulting \sys is indeed grassroots.  Section \ref{section:related-work} reviews extant work \sys is based upon, namely the Flash payment system~\cite{lewispye2023flash} and grassroots social networking~\cite{shapiro2023gsn}.  Appendix \ref{appendix:preliminaries} provides the needed mathematical preliminaries.

\section{Grassroots Cryptocurrencies}\label{section:gc}  
Grassroots cryptocurrencies~\cite{shapiro2022gc} are a permissionless digital means for turning mutual trust into liquidity.  Their coins are IOUs, units of debt that can be issued digitally by anyone---people, communities, corporations, banks, municipalities and governments, henceforth referred to as \emph{agents}---and traded by anyone. 
In this section, we give a brief overview of the way in which grassroots cryptocurrencies operate: See~\cite{shapiro2022gc} for further details.

\vspace{0.2cm} 
\noindent \textbf{The defining conventions and principles}. Grassroots cryptocurrencies have three social conventions and one principle:
\begin{enumerate}
    \item[C1.] \textbf{Minting}: Issue coins at will. 
    Each agent is the sovereign of their currency, and may voluntarily `mint' and `burn' the coins of their own currency. 
    \item[C2.]  \textbf{Pricing}: Price offerings voluntarily in terms of one's own currency.
    People and corporations can price their goods and services in terms of their own currency. 
    \item[C3.] \textbf{Mutual Credit:}  Create mutual lines of credit by voluntary coin exchange. 
    Mutual credit lines with ensuing liquidity are formed by the voluntary exchange of coins among agents with mutual trust, each providing the other with coins of their own currency, effectively exchanging IOUs.  Mutual lines of credit can thus form among family members, friends and peers;  between a community bank and its members; between a corporation and its employees, suppliers and customers;  between federal, state and municipal governments;  between a central bank and commercial and community banks; and between states.
    \item[P1.] \textbf{Coin Redemption}: Agree to redeem any domestic coin issued in return for any foreign coin held.  For example,  Alice must agree to a request by Bob to redeem an Alice-coin Bob holds against any coin held by Alice.  Bob may request from Alice in return a Bob-coin, in case Bob wishes to reduce exposure to debt by Alice, or a Charlie-coin, in case Bob needs to pay Charlie in Charlie-coins for a purchase, or wants to increase exposure to Charlie and decrease exposure to Alice.
\end{enumerate}

\mypara{Example: Autarkic Community} Consider how an isolated  community may, first, achieve liquidity and, second, establish a community bank, without any external resources such as capital or external credit.  Assume
an autarkic village has 501 productive inhabitants that all know and trust each other to some degree, each pricing their goods and services in terms of their own personal coins.
Next, assume that every two villagers exchange 100 personal coins with each other.  The result would be that each villager would have a line of credit of 50,000 coins from its fellows, each with a $0.2\%$ exposure to debt by each other villager, and that the total liquidity in the village, or coins in circulation, would be 25,000,000 coins.  Certainly something to work with.  

As long as all villagers are more or less balanced, and spend as much as they earn, with no villager overspending and no villager acting as a cash-hog, all will be well.  Still, each villager $p$ who wishes to purchase something from villager $q$ will have to use its $q$ coins, or obtain $q$-coins from others if it has run out of them.  To avoid this hassle, the village may decide, democratically, to establish a community bank.  

Assume the village has a democratically-elected board, members of which are given signature rights in a multisignature account for the village bank, and thus can issue a village coin via this account.  Assume further that the community bank opens a mutual line of credit of 1000 coins with each villager---exchanging 1000 village coins in return for 1000 of the villager's coins, and that the villagers now price their goods and services in terms of village coins, so that its economy runs even more smoothly.

One may ask what value as collateral do the 1000 villager coins the bank now holds have?  The answer is the 50,000 coins the villager has received from other villagers  (or whatever the villager has at present).  Thus, as a community bank, the only collateral the bank has are the personal IOUs of its members. However, it can shift its risk among villagers very easily.  If villager $p$ appears to the bank to be haemorrhaging money, the bank may try, while $p$ is still solvent, to redeem from $p$ some of the 1000  $p$-coins it holds against coins of other villagers held by $p$, who appear to the bank to be more financially stable.  This way, the bank can reduce its exposure to $p$ without closing down or even without reducing the 1000-coins line of credit the bank has granted  $p$ to begin with.
In effect, the bank's credit line to $p$ is now covered by $p$'s friends.

\mypara{Economic Measures}
An agent manifests lack of liquidity by  not holding coins by other agents, and not being able to obtain such coins by trade or coin exchange. Lack of liquidity may result in insolvency, making the coins/IOUs issued by the agent `bad debt', and causing those coins to be traded (on secondary markets) at a discount, if at all.

\mypara{The Role of the Sovereign} Grassroots cryptocurrencies are similar to fiat currencies issued by sovereign states in that both are collateral-free units of debt.  Hence, measures applied to fiat currencies, such as foreign debt, trade balance, and velocity,
are also relevant to assessing the financial health of a grassroots cryptocurrency.  Similarly, equivalents of the standard measures can be applied to assess the creditworthiness of an agent within a grassroots digital economy, including cash ratio, quick ratio, and current ratio~\cite{shapiro2022gc}.

If the sovereign $p$ fails to present computational or economic integrity, the result would most probably be a bank run on $p$,  insolvency, and financial harm to the family, friends, neighbours and colleagues who provided $p$ with credit.  Hence, normal normative people that participate in the grassroots economy through mutual credit lines with family, friends, neighbours and colleagues, would maintain their computational and financial integrity.

\section{The Blocklace}\label{section:blocklace}

The payment system for grassroots cryptocurrencies is implemented using the \emph{blocklace}, defined herein.

\begin{figure}[t]
  \begin{center}
   \includegraphics[width=5cm]{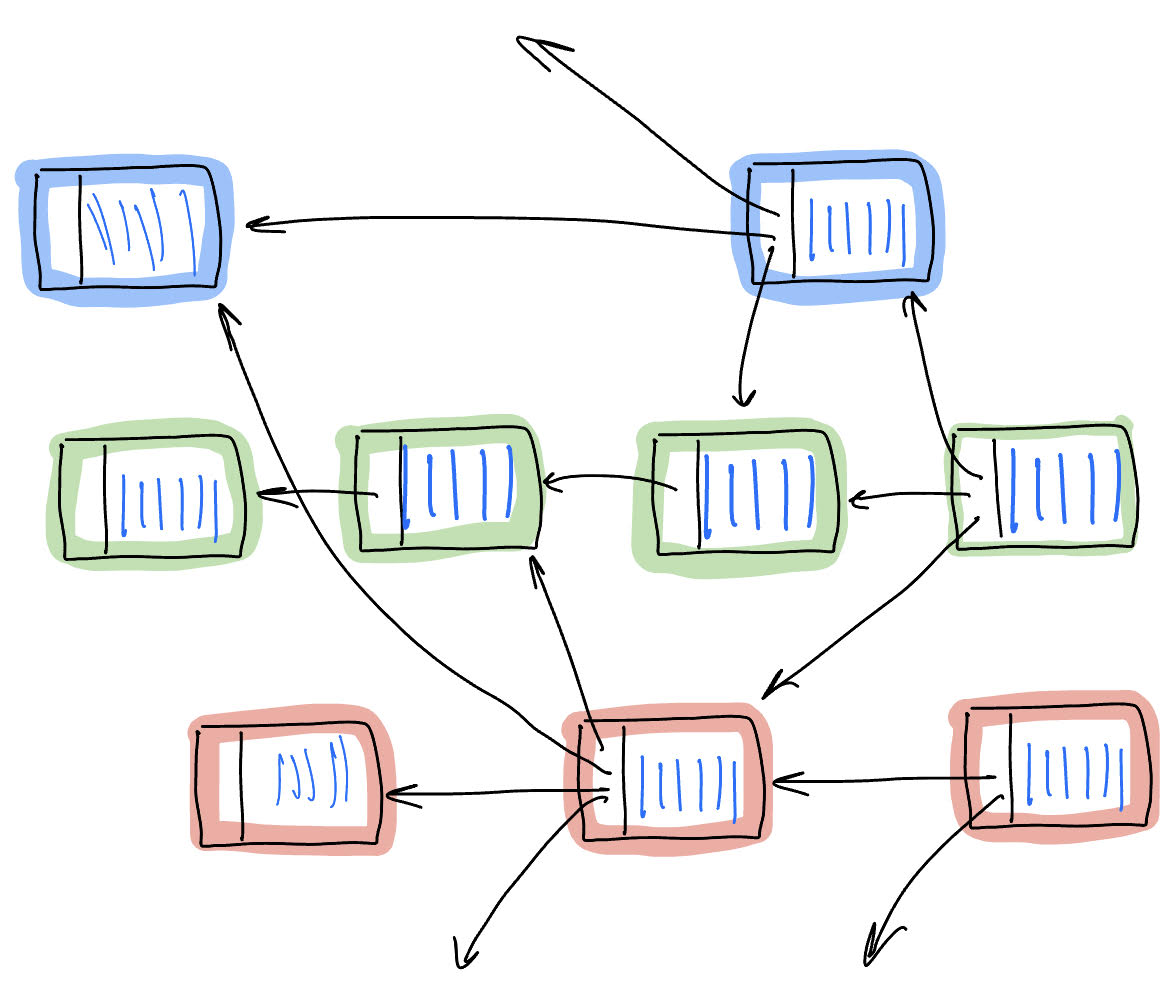}
  \end{center}
  \caption{A self-closed blocklace: Blocks are color-coded by agent, thus pointers among blocks of the same color are self-pointers, others are non-self pointers, which may be dangling.
  }
\label{figure:p-closure}
\end{figure}

\vspace{0.1cm} 
\noindent \textbf{Defining the blocklace}. The blocklace is a distributed, partially-ordered counterpart of the replicated, totally-ordered blockchain data-structure~\cite{shapiro2023grassroots} (See Figure \ref{figure:p-closure} for an example). It has already been utilized in grassroots social networking~\cite{shapiro2023gsn}, the All-to-All Flash payment system~\cite{lewispye2023flash}, and the Cordial Miners family of consensus protocols~\cite{keidar2023cordial}.  What follows below is a concise introduction to its basic concepts.

We assume a potentially-infinite set of agents $\Pi$ (think of all the agents yet to be produced), but when referring to a subset of the agents $P \subseteq\Pi$ we assume $P$ to be finite.
Each agent is associated with a single and unique key-pair of its own choosing, and is identified by its public key $p \in \Pi$.  We  also assume given a cryptographic hash function \textit{hash}. 
A \emph{block} created by agent $p \in \Pi$, also referred to as a \emph{$p$-block}, is a triple $b=(h,x,H)$, with $h$ being
a hash pointer $\textit{hash}((x,H))$ signed by $p$, also referred to as a \emph{$p$-pointer}; $x$ being referred to as the \emph{payload} of $b$; and $H$ a finite set of signed hash pointers. If $H = \emptyset$ then $b$ is a \emph{genesis block}; a hash pointer $h'\in H$ points to the block $(h',x',H')$, if such a block exists; and if $h'$ is a $p$-pointer it is also called a \emph{self-pointer}.

A \emph{blocklace} $B$ is a set of blocks, and is called \emph{closed} if every pointer in every block in $B$ points to a block in $B$.   A pointer that does not point to a block in $B$ is \emph{dangling} in $B$, hence a closed blocklace has no dangling pointers.  A blocklace is \emph{self-closed} if it has no dangling self-pointers; note that a blocklace can be self-closed but not closed, having dangling non-self pointers, as is the case for the blocklace in Figure \ref{figure:p-closure}.

The nature of a cryptographic hash function means that it is computationally infeasible to create hash pointers that form a cycle. We are concerned only with computationally feasible blocklaces, and any such blocklace $B$ induces a DAG, with the blocks of $B$ as vertices and with directed edges $b\rightarrow b'$ for every $b, b' \in B$ for which $b$ includes a pointer to  $b'$.

A block $b$ \emph{observes} a block $b'$ in $B$ if $b=b'$ or there is a path $b=b_1 \rightarrow b_2 \ldots \rightarrow b_k = b'$ 
in $B$, $k\ge 1$.  An agent $p$ \emph{observes} a block $b$ in $B$ if there is a $p$-block $b'\in B$ that observes $b$. We note that the `observes' relation among blocks is a partial order on $B$.

Two $p$-blocks that do not observe each other are referred to as an \emph{equivocation} by $p$, and if their payloads include conflicting financial transactions by $p$, they indicate an attempt by $p$ to \emph{double-spend}. If $B$ includes an equivocation by $p$ we say that $p$ is an \emph{equivocator} in $B$.  A $p$-block $b$ \emph{approves} a $q$-block $b'$ in a blocklace $B$ if $b$ observes $b'$ in $B$ and does not observe a $q$-block equivocating with $b'$ in $B$.  An agent $p$ \emph{approves} a $q$-block $b'$ in $B$ if there is a $p$-block in $B$ that approves $b'$.

A blocklace $B$ is a \emph{personal blockchain} of agent $p$, also referred to as a $p$-blockchain, if $B$ consists of $p$-blocks in which each non-genesis block has exactly one self-pointer, and $B$ is self-closed.  Note that a $p$-blockchain, while self-closed, may still contain dangling pointers to non-$p$-blocks (as in Figure \ref{figure:p-closure}), and as such it is not an `ordinary' blockchain.  If $p$ equivocates in a $p$-blockchain $B$, then $B$ must include $p$-forks (equivocating $p$-blocks pointing to the same $p$-block)  and/or multiple genesis $p$-blocks.
In the following, a correct agent does not equivocate and includes in each block a self-pointer to its previous block.
Hence, the personal blockchain of a correct agent $p$ is a \emph{single chain} -- a sequence of $p$-blocks, each with a self-pointer to its predecessor, ending in a genesis $p$-block (as is the case for all three personal blockchains in Figure \ref{figure:p-closure}).

The \emph{local state} of each correct agent $p$ is a blocklace, consisting of its personal blockchain and blocks received by $p$. A correct agent $p$ maintains a self-closed $p$-blocklace, buffering received out-of-order blocks.  
A \emph{configuration} $c$ consists of a set of local states, one for each agent.  The local state of agent $p$ in $c$ is denoted by $c_p$.

\section{\sys: Specification} \label{section:spec}

\begin{figure}[!ht]
\centering
\includegraphics[width=10cm]{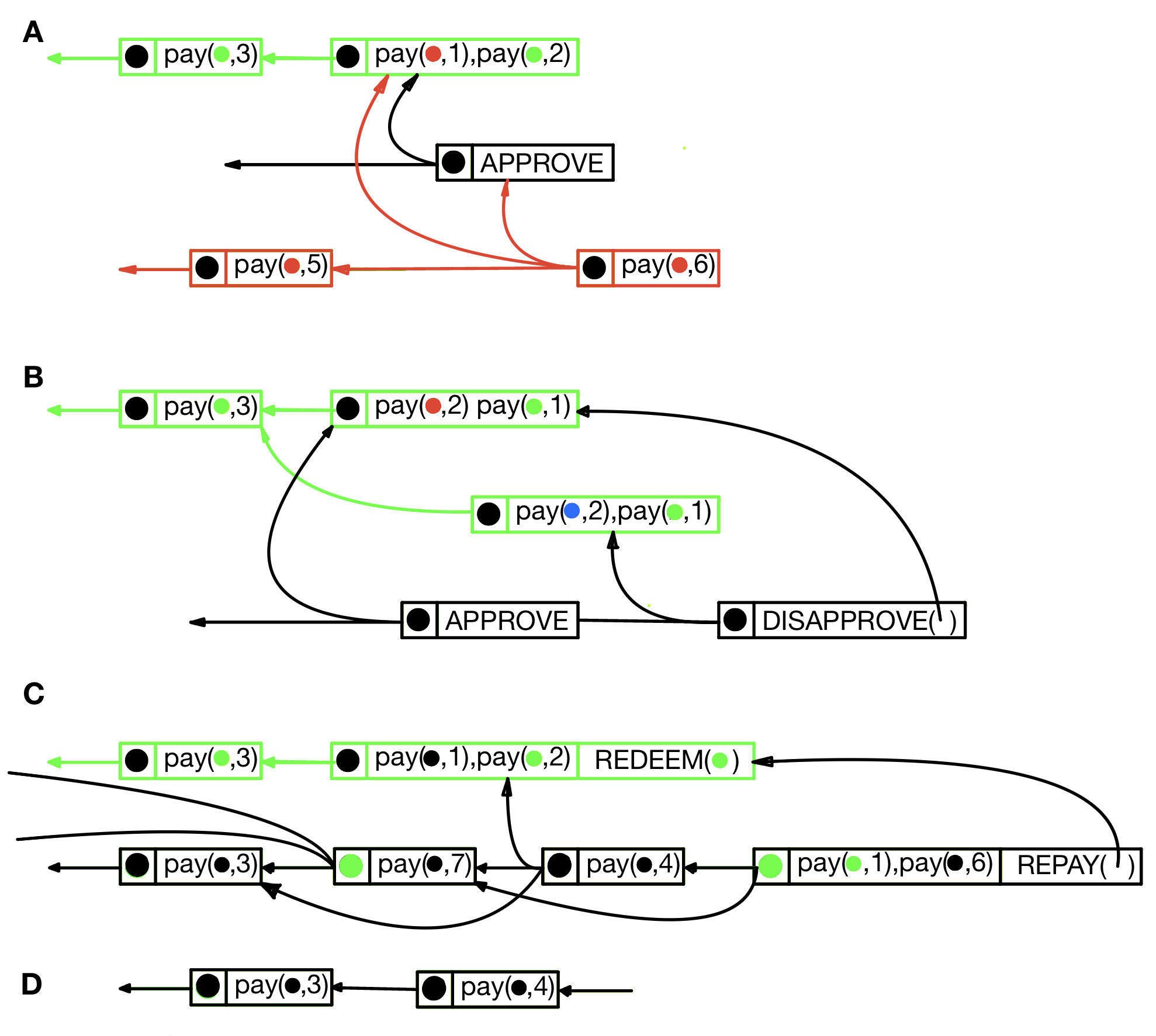}
\caption{\textbf{Example Transactions.} Agents are colored.  Box color indicates the creator of the block, leftmost dot in a block indicates the currency of the transaction.  Comments are omitted if not important.
\textbf{(A) Payment: Issue, Approve, Accept}: 
The green agent, with a balance of 3 black coins, issues a payment to the red agent of 1 black coin, decreasing its balance to 2 black coins.  The black agent approves the payment, and the red agent accepts the payment, increasing its balance to 6 black coins.
\textbf{(B) Equivocation and its Disapproval}:  The green agent equivocates, issuing two black-coins payment blocks that consume the same black-coins self-payment block.  The black agent approves the payment of the green agent to the red agent and disapproves the payment of the green agent to the blue agent, pointing to the payment to the red agent as evidence for the equivocation.
\textbf{(C) Redemption and Repayment}: The green agent redeems a black coin from the black agent, requesting a repayment of a green coin.  The black agent accepts the payment of the black coin from the green agent and repays the green agent a green coin, increasing its balance of black coins from 3 to 4 and decreasing its balance of green coins from 7 to 6.
\textbf{(D) Minting/Burning}:  The black agent mints a black coin, increasing its balance from 3 to 4 coins. 
}
\label{figure:gf}
\end{figure}

To simplify our presentation, we first specify and describe an implementation of Grassroots Flash assuming all-to-all dissemination or, equivalently, a global blocklace where each block created by a correct agent is eventually received by every correct agent.  As all-to-all dissemination is not grassroots~\cite{shapiro2023grassroots},  the task of specifying and describing a grassroots implementation is addressed in Section \ref{section:cordial-dissemination}.

\vspace{0.2cm} 
\noindent \textbf{A currency for each agent}. With each agent $r \in \Pi$ we associate an \emph{$r$-currency}, with $r$ being its \emph{sovereign}, consisting of \emph{$r$-coins}, which can be thought of as IOUs issued and signed by $r$.  Figure \ref{figure:gf} shows transactions in a black currency, with the black agent being the sovereign.
Any agent (including $r$) that issues or receives payments in $r$-coins  is referred to as a \emph{trader} in the $r$-currency. Figure \ref{figure:gf} shows transactions by the red and green traders.

\sys employs \emph{transaction blocks} with payloads of the form $x=(r,y,c)$.  Such a payload encodes a transaction in $r$-coins, with $y$ being a set of $r$-coin payments and $c$ a comment that indicates the reason for the payments.  
A \emph{payment} is a pair $(p,z)$ specifying a payment to agent $p$ of $z$ coins. 
If a payment $(p,z)$ occurs in a $p$-block it is referred to as a \emph{self-payment}.  An $r$-coin self-payment by a correct agent $p$ records the current balance of $p$ in $r$-coins. 
An $r$-coin self-payment of an agent $p$ is used to update an agent's balance in $r$-coins, for example upon receiving an $r$-coin payment or issuing an $r$-coin payment.
Figure \ref{figure:gf} shows many payment blocks, for example the top-right block is a black-coin payment block by the green agent that includes a payment of 1 black coin to the red agent and a self-payment of 2 black coins.

\begin{definition}[Consumes, Balanced, Approved/Disapproved]
Given a blocklace $B$ and two transaction blocks $b, b' \in B$ in $r$-coins where $b$ points to $b'$, we say that $b$ \temph{consumes} $b'$ if
$b$ is a $p$-block and $b'$ includes a payment to $p$.  A transaction $p$-block is \temph{balanced} if the sum of its payments and the sum of the payments to $p$ in the blocks it consumes are equal.
A payment $p$-block $b \in B$ with a payment of $r$-coins is \temph{approved} in $B$ if $p = r$ or if $p \ne r$ 
and $b$ is pointed to by an $r$-block $b'\in B$ with payload $\textsc{approve}$, 
and $b$ is \temph{disapproved} in $B$ if it is pointed to by an $r$-block $b'\in B$ with payload $(\textsc{disapprove},h')$, where $h'$ points to $b$, in case $b$ is unbalanced, or to an equivocation with $b$ in $B$. 
\end{definition}
All the payment blocks in Figure \ref{figure:gf} are balanced, except the bottom-right block that shows the black sovereign minting a new black coin.  In \ref{figure:gf}.\textbf{A} The second red block  consumes the second green block, which consumes the first green block; in \ref{figure:gf}.\textbf{B} the two equivocating green blocks consume the first green block; and in
\ref{figure:gf}.\textbf{C} the third black block consumes the first black block and the second green block, and the last black block consumes the second black block.  Figure \ref{figure:gf}.\textbf{B} also shows examples of approval and disapproval of payments by the black sovereign.

\begin{definition}[Redemption claims and their repayment]
A $p$-block $b=(h,(r,y,(\textsc{redeem},P)),H)$, with $y$ including a payment $(r,z)$ and $P$ being a nonempty list of agents, is referred to as a \temph{redemption claim} by $p$ against $r$.  
An $r$-block with a payment to $p$ that includes the comment $(\textsc{repay},h)$ is said to \temph{repay} the redemption claim $b$. 
\end{definition}
Figure \ref{figure:gf}.\textbf{C} shows a redemption of a black coin by the green agent from the black agent, and a repayment of a green coin by the black agent to the green agent.

Agent $r$ can repay the redemption claim $b$ above by paying to $p$ a total of $z$ $P$-coins, in the order of preference specified by $P$, and if it does not hold sufficiently many $P$-coins then paying any remaining balance in $r$-coins. The simplest example is when $p$ wishes to close or reduce a mutual credit line with $r$ by redeeming $r$-coins held by $p$ against $p$-coins held by $r$.  To do so, $P=[p]$, and $r$ repays the claim by paying back to $p$ up to $z$ $p$-coins, and only if $r$ does not have enough $p$-coins then it pays the remainder in $r$-coins, in effect not letting $p$ reduce the level of credit to $r$ as much as it wishes.  As another example, if  $P=[p_1,p_2]$, then $r$ pays to $p$ up to $z$ $p_1$-coins, and only if $r$ does not have enough $p_1$ coins then $r$ pays any remainder in $p_2$ coins,  and only if $r$ does not have enough $p_2$ coins then $r$ pays to $p$  any remainder in $r$-coins.

Correct agents participating in a grassroots payment protocol satisfy the following:
\begin{tcolorbox}[breakable,colback=gray!5!white,colframe=black!75!black]
\textbf{\larger Desiderata for Trader Correctness}  \newline
A correct agent $p$ trading in $r$-coins satisfies the following:
\begin{enumerate}
    \item \textbf{Safety}: 
    \begin{enumerate}
        \item \textbf{Record}: A self-payment $(p,x)$ in $r$-coins records the balance $x$ of $r$-coins held by $p$. 
        \item \textbf{Balance}: A payment $p$-block in $r$-coins is balanced (except if $p=r$ and it is a self-payment). 
        \item \textbf{Integrity (No Double-Spend)}: Each block with a payment to $p$ is consumed by at most one $p$-block.
        \item \textbf{Finality}: A $p$-block consumes a non-$p$-block $b'$ with a payment to $p$ only if $r$ approves $b'$. 
    \end{enumerate}  
    \item \textbf{Liveness}  \newline
    \textbf{Accept}: Every approved $q$-block, $q\ne p$, with a payment to $p$ is eventually consumed by a $p$-block [provided $p$, $q$ and $r$ have a common friend in $SG(B)$]\footnote{\label{a} The bracketed additions here and below apply to grassroots dissemination over the social graph $SG(B)$, as elaborated in Section \ref{section:cordial-dissemination}.   In this section, the Desiderata should be read without the bracketed additions.  In Section \ref{section:cordial-dissemination}, we refer to these Desiderata again, upon which they should be re-read with the bracketed additions.
    }.
\end{enumerate}
\end{tcolorbox}

The following notion of the balance of an agent is employed in the desiderata for sovereign agent correctness.
\begin{definition}[Balance]
Given a blocklace $B$ and two agents $p, r\in \Pi$, the $r$-coins \temph{balance} of $p$ in $B$ is the sum of $r$-coins payments accepted by $p$ in $B$ minus the sum of $r$-coins payments issued by $p$ in $B$. 
\end{definition}
\mypara{Notes} In the definition above, self-payments can either be excluded or included in both sums so they cancel each other. An incorrect agent may double-spend, or issue unbalanced payments, and reach a negative balance. An incorrect sovereign $r$ my wreck any havoc it pleases in its own currency, possibly rendering the balance in $r$-coins of any trader  meaningless.

\begin{tcolorbox}[breakable,colback=gray!5!white,colframe=black!75!black]
 \textbf{\larger Desiderata for Sovereign Agent Correctness} \newline 
A correct sovereign agent $r$ satisfies the following: 
\begin{enumerate}
    \item \textbf{Safety}
    \begin{enumerate}
    \item \textbf{Approve}: An $r$-block $b$ with an empty payment and comment $\textsc{approve}$ points to balanced $r$-coin payments block $b'$, where  $b$ does not observe a block equivocating with $b'$.
    \item \textbf{Disapprove}: An $r$-block $b$ with an empty payment and comment $(\textsc{disapprove},h')$ points to an $r$-coin payment block $b'$, where $h'$ points to the reason for disapproval: To $b'$, if it is unbalanced, or to a block $b''$ equivocating with $b'$. 

    \item \textbf{Repay}: An $r$-block $b$ that repays a redemption claim with comment $(\textsc{redeem},P)$,  $P=[p_1,p_2,\ldots,p_k]$, $k\ge 1$, has a payment in $p_i$-coins, $i \in [k]$, provided the balance of $r$ does not include any $p_j$-coins for any $1\le j < i$ at the time of creating $b$, and has a payment in $r$-coins, $r\notin P$, provided the balance of $r$ does not include any $P$-coins at the time of creating $b$. 
\end{enumerate}
    \item \textbf{Liveness} 
\begin{enumerate}
    \item \textbf{Approve/Disapprove}:  Every $p$-block with a payment in $r$-coins by a correct trader $p\ne r$ is eventually approved or disapproved by an $r$-block [provided $p$ and $r$ are friends or have a common friend in $SG(B)$]$^a$.
    \item \textbf{Repay}: Every redemption claim $b=(h,(r,x,(\textsc{redeem},P)),H)$ by a correct trader $p$ against $r$ with a payment $(r,z)$  is eventually followed by $r$-blocks with the comment $(\textsc{repay},h)$ and payments to $p$ totalling $z$ [provided $p$ and $r$ are friends in $SG(B)$]$^a$. 
\end{enumerate}
\end{enumerate}
\end{tcolorbox}

To satisfy these safety and liveness desiderata, the protocol employs the following types of blocks.  To keep the exposition simple, a $p$-block may issue at most one non-self payment,  accept one incoming payment, or approve one $p$-coins payment,  although in practice several $p$-transactions in the same currency can be aggregated in one $p$-block.

Next, we define the notion of correct blocks and naturally assume that correct agents produce only correct blocks.
\begin{tcolorbox}[breakable,colback=gray!5!white,colframe=black!75!black]
\begin{definition}[Correct Blocks] A \temph{correct} $r$-coins transaction $p$-block $b=(h,(r,y,c),H)$ includes in $H$ a pointer to the preceding $p$-block and  has one of the following four forms:
\end{definition}
\begin{enumerate}
\item \textbf{Trader Blocks} that are balanced: 
\begin{enumerate}
    \item \textbf{Issue}. A payment of $z'>0$ to $p'\ne p$ with comment $c$,  $y=((p',z'),(p,z))$, $z\ge 0$, and $H$ including a pointer to an $r$-coins self-payment $p$-block. 
    \item \textbf{Accept}. A payment  $(p,z)$ in $y$, $z > 0$,  $p\ne r$, 
    and $H$ includes a pointer to a preceding $r$-coins self-payment $p$-block, if there is one, a pointer to a non-$p$-block $b'$ with an $r$-coin payment to $p$, and a pointer to an $r$-block that approves $b'$.
    \end{enumerate}
 \item \textbf{Sovereign Agent Blocks}  with $p=r$:
\begin{enumerate}
 \setcounter{enumii}{2}
    \item \textbf{Mint/Burn  $r$-coins}. A resulting balance of $z$ $r$-coins,  $y=(r,z)$,  $H$ including a pointer to a preceding $r$-coins self-payment $r$-block, if there is one. 
    \item \textbf{Approve/Disapprove an $r$-coins payment by $q\ne r$}. \newline
    With $y=\emptyset$,  $H$ including a pointer to a $q$-block $b'$ with $r$-coins payments, $q \ne r$, and comment $c=\textsc{approve}$ if $b$ does not observe an equivocation with $b'$ and if $b'$ is balanced, else $c=(\textsc{disapprove},h'')$ if $b$ observes a $q$-block $b''$ with hash pointer $h''$ that equivocates with $b'$ or if $b'$ is unbalanced with hash $h''$.
    \end{enumerate}
\end{enumerate}
\end{tcolorbox}

\begin{proposition}[Balance Safety]\label{proposition:balance-safety}  
Given a self-closed blocklace $B$ and two  agents $p, r\in \Pi$, if $p, r$ are correct then the $p$-blocks in $B$ are totally-ordered and  the last  
$r$-coins self-payment of $p$ in $B$ equals the $r$-coins balance of $p$ in $B$.
\end{proposition}
\begin{proof}
The proof is by induction on the number $n$ of $r$-coins self-payments of $p$ in $B$.
Assume $n=0$.
Since $p$ is correct, the first $r$-coins self-payment consumes an $r$-coins payment $(p,z)$ to $p$ and includes the self-payment $(p,z)$, by definition.

Assume $n>0$ with the last $r$-coins self-payment of $p$ in $B$ being $(p,z)$ (`last' is well-defined even though $B$ is partially-ordered since $p$ is correct by assumption and hence does not equivocate in $B$).
Let $B'\supset B$ be a blocklace with $n+1$ $r$-coins self-payments of $p$, with the additional $r$-coins self-payment $p$-block $b$ including the self-payment $(p,z')$.  Since $p$ is correct by assumption, then $b$ must be a Trader block (Issue or Accept), which is balanced assumption, implying that $z'$ equals the $r$-coins balance of $p$ in $B'$.\qed 
\end{proof}

The following liveness proposition assumes that every block by a correct agent is eventually known by every correct agent.  In Section \ref{section:cordial-dissemination}, we relax this requirement and restate a weaker liveness proposition for grassroots dissemination on the social graph.

\begin{proposition}[Payment Liveness for All-to-All Dissemination]\label{proposition:payment-liveness}
Given three correct agents $p,q, r\in \Pi$,  every $r$-coins payment issued by $p$ to $q$ is eventually accepted by $q$. 
\end{proposition}
\begin{proof}
Since $p$ is correct the payment is not an equivocation; since $r$ is correct it will eventually approve the payment;  since $q$ is correct it will eventually receive the approved payment.\qed   
\end{proof}

We note that if an agent $p$ violates Trader Safety or Sovereign Agent Safety in a blocklace $B$, then this violation by $p$ can be deduced from $B$. Namely, an agent $p$ cannot make a safety violation without leaving a ``smoking gun'' in the form of $p$-blocks, signed by $p$, that witness the violation.  For example, $p$ issuing or accepting an $r$-coins payment without recording correctly via a self-payment its resulting balance in $r$-coins, results in an unbalanced $p$-block.   Equivocation by $p$ results in two equivocating $p$-blocks.  Sovereign agent $r$ approving an equivocation by $p$ results in an equivocation by $r$ or in an $r$-block that approves one of the equivocating $p$-blocks while observing the other.  

On the other hand, when there is no bound on message delay, as is the case in the model of asynchrony, liveness violations cannot be detected in finite time.  However, sovereigns that fail to respond in reasonable time would be punished by the market:  A sovereign $r$ that fails to respond to $r$-coins redemption claims, or is lazy in approving payments in $r$-coins, risks its coins being considered bad debt by anyone who is not privy to the reasons for this behavior and thus might forgive it at least temporarily.  This could result in a bank run, with agents trying to redeem their $r$-coins while it is still possible or else sell them as bad debt, at a discount.

\section{\sys with Cordial Dissemination}\label{section:cordial-dissemination}

\begin{figure}[h]
\begin{algorithm}[H]
	\caption{\textbf{\sys: The Payment System}\\ 
        \smaller Code for agent $p$}	\label{alg:gf-payment}
	\small
	\begin{algorithmic}[1] 
		  \alglinenoNew{counter}

\StateX \textbf{Global/Local Blocklace:}
\State $B \gets B \cup \{ \textit{create\_block}((p,0),\emptyset)\}$  \Comment{The genesis block of agent $p$ is added to the global blocklace/to the initially-empty local blocklace}  \label{tl:blocklace}
        \StateX \textbf{(a) Issue}
	\Upon{decision to issue a payment $(p',z')$ in $r$-coins with comment $c$}  
        \State Create a block $b$ with payload $(r,((p',z'),(p,z-z'),c)$ that consumes the previous $p$-block in $B$ with self-payment $(p,z)$ in $r$-coins, provided $z\ge z'$.
        \State $B \gets B \cup \{b\}$
        \EndUpon
 \StateX \textbf{(b) Accept}
\Upon{decision to accept a $q$-block $b' \in B$ with a payment $(p,z)$ in $r$-coins and comment $c$}  
        \State Create a block $b$ with payload $(r,(p,z+z'),c)$ that consumes $b'$ and the previous $p$-block in $B$ with self-payment $(p,z')$ in $r$-coins, if there is one, else $z'=0$.
        \State $B \gets B \cup \{b\}$
        \If{$p=r$ and $c=(\textsc{redeem},P)$} issue repayments to $q$ in $P$-coins if possible and in order, with the remainder in $r$-coins, totalling $z$.
        \EndIf
 \EndUpon
 \StateX \textbf{(c) Mint/Burn} with $p=r$
\Upon{decision to mint/burn $r$-coins with resulting balance $z$ and comment $c$}  
        \State Create a block $b$ with payload $(r,(r,z),c)$ that consumes the previous $r$-block in $B$ with self-payment, if there is one.
         \State $B \gets B \cup \{b\}$
        \EndUpon
 \StateX \textbf{(d) Approve/Disapprove} with $p=r$
\Upon{decision to approve/disapprove block $b \in B$ with comment $c$}  
        \State Create a block $b$ with payload $(r,\emptyset,c)$ that points to $b$ with $c=\textsc{approve}$ or $c=(\textsc{disapprove},h')$, as the case may be.
         \State $B \gets B \cup \{b\}$
        \EndUpon

    \alglinenoPush{counter}
\end{algorithmic}
\end{algorithm}
\vspace{-2em}
\end{figure}

We recall (Observation 2 in~\cite{shapiro2023grassrootsBA}) that an all-to-all dissemination protocol cannot be grassroots.
Intuitively, if a dissemination protocol is grassroots then a group of agents engaged in it may ignore additional agents, in contradiction to the dissemination being all-to-all.  Practically, a digital economy in which every transaction has to be disseminated to every member of the economy is not scalable.

Hence, in this section, we replace the assumption of all-to-all dissemination with Cordial Dissemination, argue that the change does not affect safety, prove a liveness proposition (Proposition \ref{proposition:CD-liveness}) and prove that the resulting Grassroots Flash protocol with Cordial Dissemination is indeed grassroots (Theorem \ref{theorem:gf-grassroots}).

We refer the reader to Appendix \ref{appendix:preliminaries}, which recalls the necessary mathematical background~\cite{shapiro2021multiagent,shapiro2023grassroots}, and introduce the notion of a grassroots protocol without further ado. In the definition below, $TS(P)$
denotes the set of \emph{all possible behaviours} of a set of correct agents $P$ that follow the protocol $TS$, while a group of agents $P$  has possible behaviors $TS(P')/P$ when embedded within a larger group $P'$ (see Appendix \ref{appendix:preliminaries} for details). 

\begin{definition}[Grassroots]\label{definition:grassroots}
A  protocol $\calF$ is \temph{grassroots} if $\emptyset \subset P \subset P' \subseteq \Pi \text{ implies that }\\ TS(P) \subset TS(P')/P$. 
\end{definition}

  A protocol is grassroots if: (\ia) The behaviors of the agents $P$ on their own,  $TS(P)$,
are also possible behaviors of these agents when embedded within a larger group $P'$, $TS(P')/P$. In other words, the agents in $P$ may choose to ignore the agents in $P'\setminus P$.  Hence the \textbf{subset relation}.  (\ib) This latter set of behaviors $TS(P')/P$ includes additional possible behaviors of $P$ not in $TS(P)$.  Thus, there are possible behaviors of $P$, when embedded within $P'$, which are not possible when $P$ are on their own.  This is presumably due to interactions between members of $P$ and members of $P'\setminus P$.  Hence the subset relation is \textbf{strict}.

To prove that the Grassroots Flash protocol is indeed grassroots, we will employ the
following Theorem (originally Theorem 12 of reference~\cite{shapiro2023grassroots}):
\begin{restatable}[Grassroots Protocol]{theorem}{GrassrootsProtocol}\label{theorem:grassroots}
An asynchronous, interactive, and non-interfering protocol is grassroots.
\end{restatable}

Informally, a protocol is \emph{asynchronous} if a  transition by an agent, once enabled, cannot be disabled by transitions taken by other agents; it is \emph{interactive} if the addition of agents to a group results in additional possible behaviors of the group; and it is \emph{non-interfering} if the possible behaviors of a group of agents are not hampered by the presence of additional stationary agents, namely agents that remain in their initial state during the entire computation.  See Apendix \ref{appendix:preliminaries} for a definition.

\mypara{Cordial Dissemination} 
In Cordial Dissemination~\cite{shapiro2023grassroots} agents create blocks, store (\emph{know}) blocks by them and others, and express \emph{needs} to know blocks by others.
Agents may become \emph{friends} by mutual consent (and break a friendship unilaterally), and upon doing so may communicate.
 
The friendship relation among the agents induces an undirected graph with agents as vertices and friendships as edges, referred to as the \emph{social graph}. The social graph is distributed,  in the sense that each agent stores only part of the graph---its local neighbourhood---and is \emph{decentralized}, in the sense that each agent has control over its local neighbourhood: It can form a new edge at will, with the consent of its new neighbour, and remove an edge at will.  
Friends may communicate, with an edge in the social graph representing a communication link of
the weakest type: A fair-loss link~\cite{cachin2011introduction}, in which a message that keeps being retransmitted is eventually delivered, if both agents are correct.

Cordial dissemination employs these rather abstract notions using the following principle:
\begin{tcolorbox}[colback=gray!5!white,colframe=black!75!black]
\textbf{The Principle of \CD over a Social Graph:}\newline
Every so often, friends send each other everything they know that the other needs.
\end{tcolorbox}
A slightly more precise but cumbersome version of the statement above would be that among correct friends $p$ and $q$, every so often $p$ sends  $q$  every block $b$ such that: (\ia) $p$ knows $b$ and (\ib) $p$ knows that $q$ needs $b$ and (\ic) $p$ does not know that $q$ already knows $b$.  

We say that two friends are \emph{indefinite} if they never break their friendship.
These notions and principles have the following liveness consequence:
\begin{observation}[Liveness of Indefinite Friends]\label{observation:friends-liveness}
For two indefinite friends $p, q$, everything that $p$ eventually knows and $q$ eventually needs will be known to $q$, eventually.
\end{observation}
In the following, we assume all friendships are indefinite.

\mypara{Blocklace Realization of Cordial Dissemination} The principles of \CD over a Social Graph are realized by agents creating, sharing and storing blocks in a blocklace, as follows.

Two agents $p, q \in \Pi$ become friends simply by both creating a block stating so, a $p$-block with payload $(\textsc{friend},q)$ and a $q$-block with payload $(\textsc{friend},p)$.  For simplicity of exposition, we defer the handling of breaking a friendship.

The principle  of \CD is realized by the blocklace as follows:
Agents creating blocks with payloads.
A correct agent $p$ says what it \emph{needs} via its personal blockchain.
For friends $p,q \in \Pi$, every so often $p$ sends $q$ the blocks $p$ knows that $q$ needs but does not know, based on the portion of $q$'s personal blockchain known to $p$.  
The notion of \emph{need} is application specific (see Section \ref{section:gsn} for a review).
For \sys, we assume that friends need to know everything their friends say, as well as specific blocks by others (payment issued to them, payments issued in their coin, and approvals/disapprovals of payments to/by them), as defined below.
Reference~\cite{shapiro2023grassroots} describes a Twitter-like cordial dissemination protocol as a family of distributed multiagent transition systems and proves it to be grassroots. 

The safety assurance of the protocol is that the local blocklace of any correct agent $p$ is self-closed and has no equivocations by $p$ (but may include equivocation by faulty agents). 

\begin{figure}[h]
\begin{algorithm}[H]
	\caption{\textbf{Cordial Dissemination}\\ Code for agent $p$
 }	\label{alg:CD}
	\small
	\begin{algorithmic}[1] \scalefont{0.93}
	\alglinenoPop{counter} 
		
\Statex \textbf{Local variables:}
\State $B \gets \{ \textit{create\_block}(\bot,\emptyset)\}$  \Comment{The local blocklace of agent $p$}  \label{tl:blocklace}

\vspace{0.5em}
	\Upon{decision to create block with payload $x$}  \label{alg:create}
        \State  $b\gets \textit{create\_block}(x,\textit{tips}(B))$       \Comment{1. \textbf{Say $x$}}
         \State $B\gets B \cup \{b\}$  
	 \EndUpon
	   	 \vspace{0.5em}	
        \Upon{$\textbf{receive } b$ s.t. $B \cup \{b\}$ is self-closed and $p$ needs $b$ }   \label{alg:receive}  \Comment{3. \textbf{Receive $b$}}
	    \State $B\gets B \cup \{b\}$  
        \EndUpon
 \vspace{0.5em}	

     \Upon{every so often, for every friend $q$}
      \label{alg:a-disseminate-timeout}
	            \Comment{2. \textbf{Meet $q$}}
        \For{all $b\in B:  \textit{needs}(q, b) \wedge \lnot\textit{agentObserves}(q,b)$}   \label{alg:disseminate}
	    \State \textbf{send}  $b$ to  $q$  \label{alg:a-send-package}
	    \EndFor
	\EndUpon

\vspace{0.5em}

		\alglinenoPush{counter}
	\end{algorithmic}

\end{algorithm}
\vspace{-2em}
\end{figure}

Algorithm \ref{alg:CD} presents pseudocode implementation of  Cordial Dissemination for a single agent $p$. While presented abstractly, its intended application is via  mobile (address-hopping) agents communicating over an unreliable network, namely smartphones communicating via UDP.
Cordial dissemination over UDP exploits the ack/nak information of blocklace blocks to its fullest, by $p$ retransmitting to every friend $q$ every block $b$ that $p$ knows (not only $p$-blocks) and believes that $q$ needs, until $q$ acknowledges knowing $b$.  While in the current architecture of the Internet,  smartphone address discovery and peer-to-peer communication among smartphones are far from  trivial, all that the protocol assumes is that friends can always eventually find a way to communication, for example by establishing  a peer-to-peer TCP connection or using peer-to-peer Bluetooth communication when they are in sufficiently-close proximity.

\mypara{Cordial Dissemination for \sys}
First, we recall the safety and liveness desiderata presented above for traders and sovereign agents, but this time with the addition of the bracketed text.  We note that as Algorithm \ref{alg:gf-payment} is retained,  the safety requirements are all satisfied.  The change discussed here is replacing all-to-all dissemination with Cordial Dissemination, which affects only liveness.

To characterize the liveness properties of \sys with Cordial Dissemination, we specify a notion of \emph{needs} tailored to \sys.  First, we recall that friends need each other's blocks, so among friends, grassroots dissemination functions like all-to-all dissemination.   For example, if an $r$-coin payment is issued by $p$ to $q$, when $p$, $q$ and $r$ are all friends with each other, then the payment $p$-block, the approval $r$-block, and the accepting $q$-block will all be known to all.

Next, we show that such a transaction can be completed even if only one of the three is a friend of the other two, or even if none of the three are friends with each other, but they all share a common friend.  For example, assume that $r$ is a community bank and $p$ is a bank member (and thus a friend of the bank).  Then $p$ can pay its friend $q$ with $r$-coins even if $q$ is not not a bank member; $p$ can pay the bank-member $q'$ $r$-coins even if $p$ and $q'$ are not friends; and the friends $p'$ and $q'$ can pay each other with $r$-coins even if neither is a bank member,  provided they have a common friend $p$ who is.

To show this claim,  we need to specify the notion of need for \sys.
\begin{definition}[\sys Needs]
In \sys, every agent \temph{needs} every block issued by its friends. In addition, when $p$ issues a block $b$ with an $r$-coin payment to $q$:
\begin{enumerate}
    \item $r$ and $q$ \temph{need} $b$
    \item $p$ and $q$ \temph{need} any $r$-block that approves or disapproves $b$.
\end{enumerate}
\end{definition}

\begin{figure}[t]
  \begin{center}
   \includegraphics[width=7cm]{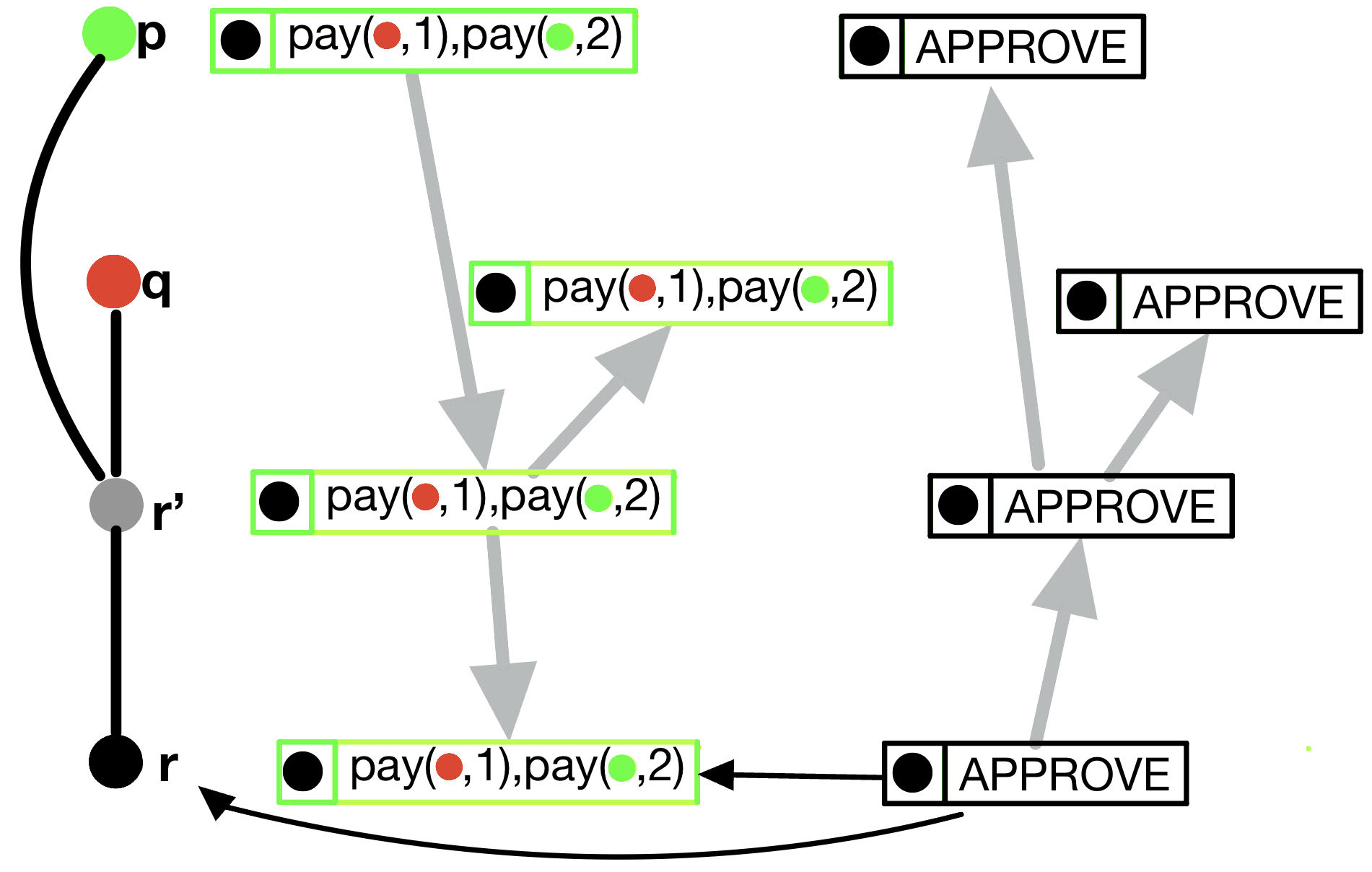}
  \end{center}
  \caption{Cordial Dissemination among agents $p$ (green), $q$ (red), $r'$ (grey) and $r$ (black).  Social graph (black lines) on the left, block dissemination (grey arrows) on the right.  A $p$-block with a payment of 1 $r$-coin to $q$ is send by $q$ to its friend $r'$, who sees that both of its friends $q$ (the recipient) and $r$ (the sovereign)  need this block, and therefore $r'$ sends it to both.  Upon receipt, the sovereign $r$ issues an approval block, which is sent to its friend $r'$. The agent $r'$ sees that both $p$ (the issuer of the payment) and $q$ (its recipient) need this approval block, and therefore sends it to both.
  }
\label{figure:gs-liveness}
\end{figure}
\begin{proposition}[Liveness of Diameter-2 Payments]\label{proposition:CD-liveness}
With Cordial Dissemination, an $r$-coin payment issued by $p$ to $q$ will eventually be accepted by $q$ if $p$, $q$, and $r$ have a friend $r'$ in common and all four agents are correct.
\end{proposition}
The following proof also holds in the special case that any one of the three parties to the transaction is a friend of the other two, formally---if $p=r'$, $q=r'$, or $r=r'$.
So for example $p$ can pay its friend $q$ with an $r$-coin if  $p$  or $q$ (but not necessarily both) is also a friend of $r$.  The general case is illustrated in Figure \ref{figure:gs-liveness}
\begin{proof}
We argue that to complete such a payment, namely for $q$ to accept the $r$-coin payment $b$ issued by $p$, it is enough that $p$, $q$, and $r$ have a common friend $r'$, provided all four agents are correct.  

Since $r'$ is a friend of $p$, it will eventually know $b$. Since $b$ includes a payment to $q$, $r'$ will  know that $q$ needs $b$ upon receipt of $b$.  Since $r'$ and $q$ are friends, $q$ will eventually know $b$.
Since $b$ is a payment in $r$-coins, $r'$ will know that $r$ needs $b$ upon receipt of $b$, and since $r'$ and $r$ are friends then $r$ will eventually know $b$.  Since $r$ is correct it will eventually issue an approve or disapprove block $b'$ for the payment block $b$, which will be known eventually to its friend $r'$.
And since $p$ is correct then in fact $b'$  approves $b$.  Since $b$ is a payment from $p$ to $q$, $b'$ is needed by both.  Since both are friends of $r'$, both will eventually know $b'$, and since $q$ is correct it will eventually accept the payment $b$.\qed
\end{proof}

Finally, we prove that the resulting \sys protocol is grassroots.

\begin{theorem}\label{theorem:gf-grassroots}
The protocol \sys is grassroots.   
\end{theorem}
\begin{proof}
The proof is structurally-identical to the proof that Twitter-like Cordial Dissemination is grassroots (Theorem 3 in reference~\cite{shapiro2023grassroots}), except that there the protocol is defined formally as a family of asynchronous multiagent transition systems, whereas here the protocol is described less formally, as pseudocode satisfying certain safety and liveness desiderata.  Hence, the argument here is accordingly less formal.
According to Theorem \ref{theorem:grassroots},
an asynchronous, interactive non-interfering protocol is grassroots.  
Consider  $\emptyset \subset P \subset P' \subseteq \Pi$. 
We argue that the \sys protocol is:
\begin{enumerate}[leftmargin=*]
    \item \textbf{Asynchronous}. Asynchrony requires that an enabled transition of an agent $p$ remain enabled despite a change of state by agents other than $p$.  The transitions of Algorithm \ref{alg:gf-payment} for agent $p$ are all based on decisions by $p$, which cannot be prevented  by other agents changing their state. In Algorithm \ref{alg:CD}, the decision to create a block (\cref{alg:create}) depends only on the internal state of $p$.  The conditions for receiving a block (\cref{alg:receive}) are monotonic, in that once they hold they hold indefinitely, so other agents cannot disable the transition once enabled.
    The seemingly non-monotonic condition $\lnot\textit{agentObserves}(q,b)$ (\cref{alg:disseminate}) refers to the local state of $p$ and hence does not hamper asynchrony.  Hence \sys is asynchronous.
    
    \item \textbf{Interactive}. Consider agents $p \in P$ and $q \in P'\setminus P$.
    In a \sys run with $P'$, $p$ may receive a payment from $q$, a transition not available in a run with $P$.  Hence 
    \sys is interactive according to Definition \ref{definition:non-interfering}.
    
    \item \textbf{Non-interfering}.  According to Definition \ref{definition:non-interfering}, we have to argue, for every $\emptyset \subset P \subset P' \subseteq \Pi$: 
\begin{enumerate}
    \item \textbf{Safety}: That every transition of agent $p\in P$ of \sys over $P$, can be carried out by $p$ when \sys is over $P'$, when agents in $P'\setminus P$ do nothing. Inspecting the transitions of $p$ in a state reachable by a run of \sys over $P$, they could only depend on blocks from agents in $P$.  Hence, one can construct a run of \sys over $P'$ in which agents in $P$ act as in the original run and agents in $P'\setminus P$ do nothing.  In the reconstructed run $p$ would reach the same state as in the original run and hence could take the same transition. 
    
    \item \textbf{Liveness}: That for every agent $p \in P$ and a run of \sys over $P$ that satisfies the liveness desiderata, a run of \sys over $P'$, when agents in $P'\setminus P$ do nothing, also satisfies the liveness desiderata.  Inspecting the liveness desiderata for Traders and Sovereign Agents in a run of \sys over $P$, they refer only to traders in $P$, hence agents in $P'\setminus P$ doing nothing do not hamper liveness of the run of \sys over $P'$, in which agents in $P$ act as in the original run and agents in $P'\setminus P$ do nothing.
\end{enumerate}
\end{enumerate} 
This completes the proof.\qed
\end{proof}

\mypara{Privacy} Friends disclose to each other their personal blockchains so, on the face of it, there can be no privacy among friends.  Such disclosure is needed for Cordial Dissemination to be fault resilient, and for friends to verify each others' computational and economic integrity.
This ability is essential, as the source of liquidity in the economy of grassroots cryptocurrencies is mutual credit lines among friends---people that trust each other, and therefore should be able to verify each other's integrity.  Still, the person behind agent $p$ may retain some financial privacy by obfuscating some of their financial transactions, as follows:  In addition to the agent $p$, the person can create a proxy $p'$ with a different keypair.  The agent $p'$ will not issue its own coins, for two reasons:  First, the collateral provided by a person for its debt would have to be split among two agents, $p$ and $p'$, to the detriment of the person's creditors.  Second, this will require disclosing the personal blockchain of $p'$ to its creditors, undermining the goal of privacy. Hence, the person will use $p'$ simply to have financial transactions one-step removed from $p$; this way the friends/creditors of $p$ will know of transfers of $p$ to $p'$, but will not know the ultimate destination of these transfers.  They may also not know of transfers to $p'$, which, in principle, should serve as collateral for the coins of $p$.  Thus, for the person behind $p'$ to be clean-handed, $p'$ should only serve as a privacy-enhancing proxy for transfers from and to $p$, and when idle have a zero balance.

\section{Relation to Extant Work}\label{section:related-work}
\sys incorporates ideas from the Flash payment system and from grassroots social networking.
\subsection{The Flash Payment System}

Flash~\cite{lewispye2023flash} (referred to here as All-to-All Flash, to distinguish it from Grassroots Flash) is a consensus-free, supermajority-based and UTXO-based payment system for the model of asynchrony. It is the first payment system to achieve $O(n)$ communication complexity per payment in the good case, when the number of Byzantine agents is constant and the network is synchronous, and $O(n^2)$ complexity in the worst-case, matching the lower bound~\cite{naor2022payment}.  It does so by sidestepping Reliable Broadcast~\cite{bracha1987asynchronous} and instead using the blocklace for the tasks of recording transaction dependencies, block dissemination, and equivocation exclusion, which in turn prevents doublespending.

All-to-All Flash  has two variants: for high congestion when multiple blocks that contain multiple payments are issued concurrently; and for low congestion when payments are infrequent.  Grassroots Flash adapts the low-congestion variant of All-to-All Flash to implement grassroots cryptocurrencies.  All-to-All Flash is not grassroots as it employs all-to-all dissemination (Observation 2 in~\cite{shapiro2023grassrootsBA}), and neither are  other payment systems that employ the all-to-all Reliable Broadcast protocol. The following table 
compares All-to-All Flash to Grassroots Flash:
\begin{center}
\smaller
\begin{tabular}{ | m{12em} | m{11em}| m{15em} | } 
    \hline
    \textbf{Protocol} & \textbf{All-to-All Flash}
 &  \textbf{\sys}
\\
     \hline
\hline
     \textbf{Cryptocurrency}  & Single & Multiple, mutually-pegged
\\
    \hline
    \textbf{Cryptocurrency supply}  & Fixed & Issued at will by sovereign
\\
     \hline
    \textbf{Permission} & Permissioned & Permissionless \\
     \hline
        \textbf{Dissemination} & All-to-All & Grassroots \\
     \hline
        \textbf{Data structure} & Blocklace & Blocklace \\
     \hline
      \textbf{Consensus-based} & No & No \\
     \hline
    \textbf{Finality by/Trust in} & Supermajority of agents
     & Sovereign (= leader)\\
 \hline
    \textbf{Transactions} &  Pay, Ack &  Pay, Redeem, Approve/Disapprove
\\
 \hline
    \textbf{Payment model}  & UTXO & UTXO
\\
    \hline
    \textbf{Complexity}  & Good-case: $O(n)$ \newline  Worst case: $O(n^2)$ & Per agent, linear in the number of their friends 
\\
 \hline   
  \end{tabular}
  \end{center}

\subsection{Grassroots Social Networking}\label{section:gsn}

Grassroots social networking~\cite{shapiro2023gsn} offers an alternative architecture to centrally-controlled global digital platforms for social networking.  
The architecture is geared for roaming (address-changing) agents communicating over an unreliable network, e.g., smartphones communicating via UDP, and realizes this principles using the following components:
\begin{enumerate}[leftmargin=*]
    \item \textbf{Decentralized Social Graph},
    with each member controlling, maintaining and storing its local neighbourhood in the graph.  
    \item \textbf{Member-Created Feeds}, each with authors and followers.
    \item \textbf{Cordial Dissemination}~\cite{shapiro2023grassroots}, variants of the protocol used here, which also use the blocklace data structure~\cite{shapiro2023grassroots,keidar2023cordial,lewispye2023flash}.
\end{enumerate}
Example grassroots social networking protocols provided include Twitter-like, with personal public feeds and followers who may comment on posts,  and WhatsApp-like, with personally created and managed private messaging groups.  The following table 
compares the use of the social graph in these two protocols to its use in Grassroots Flash:
  \begin{center}
  \smaller
 \begin{tabular}{ | m{9em} | m{9em}| m{10em} | m{13em} | } 
    \hline
    \textbf{Protocol/\newline Social Graph} & \textbf{Grassroots TL} \textbf{(Twitter-Like)}
 & \textbf{Grassroots WL (WhatsApp-Like)}  & \textbf{Grassroots Flash}
\\
     \hline
    \hline
    \textbf{Graph type} & Directed  & Hypergraph 
                                                    & Undirected  \\
     \hline
    \textbf{Edge meaning} & Source agent follows destination agent
     & A group of agents  & Mutual disclosure of personal blockchains\\
 \hline
    \textbf{Friendship \newline condition} & Mutual following & Membership in same group & Mutual friendship offers
\\
    \hline

    \textbf{Graph changes by}  & Agent follows/unfollows another agent & Agent founds group;  invites/removes agents & Agents make/break friendships
\\
 \hline 
  \textbf{Agent needs}  &  Feeds of agents it follows &  Feeds of groups it is a member of
   &  Payments to it/in its coin; approvals of payments by/to it.\\
    \hline 
  \textbf{Data Structure }  &  Blocklace &  Blocklace   &  Blocklace
\\
 \hline   
  \end{tabular}
  \end{center}

\bibliographystyle{splncs04}
\bibliography{bib}

\begin{thebibliography}{10}
\providecommand{\url}[1]{\texttt{#1}}
\providecommand{\urlprefix}{URL }
\providecommand{\doi}[1]{https://doi.org/#1}

\bibitem{agarwal2017banking}
Agarwal, S., Alok, S., Ghosh, P., Ghosh, S., Piskorski, T., Seru, A.: Banking
  the unbanked: What do 255 million new bank accounts reveal about financial
  access? Columbia Business School Research Paper (17-12) (2017)

\bibitem{auvolat2020money}
Auvolat, A., Frey, D., Raynal, M., Ta{\"\i}ani, F.: Money transfer made simple:
  a specification, a generic algorithm, and its proof. arXiv preprint
  arXiv:2006.12276  (2020)

\bibitem{benet2014ipfs}
Benet, J.: Ipfs-content addressed, versioned, p2p file system. arXiv preprint
  arXiv:1407.3561  (2014)

\bibitem{bracha1987asynchronous}
Bracha, G.: Asynchronous {B}yzantine agreement protocols. Information and
  Computation  \textbf{75}(2),  130--143 (1987)

\bibitem{bruhn2009economic}
Bruhn, M., Love, I.: The economic impact of banking the unbanked: evidence from
  mexico. World bank policy research working paper (4981) (2009)

\bibitem{cachin2011introduction}
Cachin, C., Guerraoui, R., Rodrigues, L.: Introduction to reliable and secure
  distributed programming. Springer Science \& Business Media (2011)

\bibitem{castro1999practical}
Castro, M., Liskov, B., et~al.: Practical byzantine fault tolerance. In: OsDI.
  vol.~99, pp. 173--186 (1999)

\bibitem{chockler2007constructing}
Chockler, G., Melamed, R., Tock, Y., Vitenberg, R.: Constructing scalable
  overlays for pub-sub with many topics. In: Proceedings of the twenty-sixth
  annual ACM symposium on Principles of distributed computing. pp. 109--118
  (2007)

\bibitem{collins2020online}
Collins, D., Guerraoui, R., Komatovic, J., Kuznetsov, P., Monti, M., Pavlovic,
  M., Pignolet, Y.A., Seredinschi, D.A., Tonkikh, A., Xygkis, A.: Online
  payments by merely broadcasting messages. In: 2020 50th Annual IEEE/IFIP
  International Conference on Dependable Systems and Networks (DSN). pp.
  26--38. IEEE (2020)

\bibitem{activitypub}
{W}orld {W}ide~{W}eb {C}onsortium: {A}ctivity{P}ub.
  \url{https://activitypub.rocks/}, accessed: 2023-09-10

\bibitem{dupas2018banking}
Dupas, P., Karlan, D., Robinson, J., Ubfal, D.: Banking the unbanked? evidence
  from three countries. American Economic Journal: Applied Economics
  \textbf{10}(2),  257--297 (2018)

\bibitem{fediverse}
Fediverse: {F}diverse. \url{https://fediverse.party/}, accessed: 2023-09-10

\bibitem{guerraoui2019consensus}
Guerraoui, R., Kuznetsov, P., Monti, M., Pavlovi{\v{c}}, M., Seredinschi, D.A.:
  The consensus number of a cryptocurrency. In: Proceedings of the 2019 ACM
  Symposium on Principles of Distributed Computing. pp. 307--316 (2019)

\bibitem{keidar2021need}
Keidar, I., Kokoris-Kogias, E., Naor, O., Spiegelman, A.: All you need is {DAG}
  (2021)

\bibitem{keidar2023cordial}
Keidar, I., Naor, O., Shapiro, E.: Cordial miners: A family of simple and
  efficient consensus protocols for every eventuality. In: 37th International
  Symposium on Distributed Computing (DISC 2023). LIPICS (2023)

\bibitem{lewispye2023flash}
Lewis-Pye, A., Naor, O., Shapiro, E.: Flash: An asynchronous payment system
  with good-case linear communication complexity. arXiv preprint
  arXiv:2305.03567  (2023)

\bibitem{naor2022payment}
Naor, O., Keidar, I.: On payment channels in asynchronous money transfer
  systems. In: 36th International Symposium on Distributed Computing (DISC
  2022). Schloss Dagstuhl-Leibniz-Zentrum f{\"u}r Informatik (2022)

\bibitem{shapiro2021multiagent}
Shapiro, E.: Multiagent transition systems: Protocol-stack mathematics for
  distributed computing. arXiv preprint arXiv:2112.13650  (2021)

\bibitem{shapiro2022gc}
Shapiro, E.: Grassroots cryptocurrencies: A foundation for a grassroots digital
  economy. arXiv preprint arXiv:2202.05619  (2022)

\bibitem{shapiro2023grassroots}
Shapiro, E.: Grassroots distributed systems: Concept, examples, implementation
  and applications. arXiv preprint arXiv:2301.04391  (2023)

\bibitem{shapiro2023grassrootsBA}
Shapiro, E.: Grassroots distributed systems: Concept, examples, implementation
  and applications (brief announcement). In: 37th International Symposium on
  Distributed Computing (DISC 2023). LIPICS (2023)

\bibitem{shapiro2023gsn}
Shapiro, E.: Grassroots social networking: Serverless, permissionless protocols
  for twitter/linkedin/whatsapp. In: OASIS ’23. Association for Computing
  Machinery (2023). \doi{10.1145/3599696.3612898}

\bibitem{stoica2003chord}
Stoica, I., Morris, R., Liben-Nowell, D., Karger, D.R., Kaashoek, M.F., Dabek,
  F., Balakrishnan, H.: Chord: a scalable peer-to-peer lookup protocol for
  internet applications. IEEE/ACM Transactions on networking  \textbf{11}(1),
  17--32 (2003)

\bibitem{yin2019hotstuff}
Yin, M., Malkhi, D., Reiter, M.K., Gueta, G.G., Abraham, I.: Hotstuff: Bft
  consensus with linearity and responsiveness. In: Proceedings of the 2019 ACM
  Symposium on Principles of Distributed Computing. pp. 347--356 (2019)

\end{thebibliography}
\clearpage

\appendix

\section{Preliminaries}\label{appendix:preliminaries}
\subsection{Asynchronous distributed multiagent transition systems} 
We formalize the notion of a grassroots protocol using asynchronous
distributed multiagent transition system~\cite{shapiro2021multiagent}.    Multiagent transition systems allow the description of faults, including safety (Byzantine) and liveness (fail-stop/network/Byzantine) faults and fault-resilient implementations,
and allow proving the resilience of specific protocols to specific faults, e.g. having less than one third of the agents be Byzantine and having the adversary control network delay in asynchrony.  In addition, any implementation among transition systems can be proven to be resilient to specific faults.  Thus, any implementation, including grassroots ones, can be proven resilient to specific faults
See reference~\cite{shapiro2021multiagent} for a full description introduction; we repeat below essential definitions and results.

\begin{definition}[Local States, $\prec$, Initial State]
A \temph{local states function} $S$ maps every set of agents $P \subseteq \Pi$ to a set of local states $S(P)$, satisfying
$S(\emptyset)=\{s0\}$, with $s0$ referred to as the \temph{initial local state}, and 
$P \subset P' \implies S(P) \subset S(P')$. The local states function $S$ also has an associated partial order $\prec_{S}$ on $S(\Pi)$,  abbreviated $\prec$  when $S$ is clear from the context,
with $s0$ as its unique  least element.
\end{definition}

Note that a member of $S(P)$ is a possible local state of an agent in $P$. Intuitively, think of $S(P)$ as the set of all possible sequences of messages among members of $P$, or the set of all possible  blockchains created and signed by members of $P$,  both with $\prec_S$ being the prefix relation and $s0$ being the empty sequence; or the set of all possible sets of posts/tweets and threads of responses to them by members of $P$, with $\prec_S$ being the subset relation, and $s0$ being the empty set. 

In the following, we assume a given fixed local states function $S$ with its associated ordering $\preceq_S$ and initial state $s0$.
\begin{definition}[Configuration, Transition, $p$-Transition, $\prec$]\label{definition:transition}
Given a set $X \subseteq S(\Pi)$ that includes $s0$ and a finite set of agents $P\subseteq \Pi$, a \temph{configuration} $c$ over $P$ and $X$ is a member of $C:=X^P$, namely $c$ consists of a set of local states in $X$ indexed by $P$.  
Given a configuration $c \in C$ and an agent $p \in P$, $c_p$ denotes the element of $c$ indexed by $p$, referred to as the \temph{local state of $p$ in $c$},
and $c0:=\{s0\}^P$ denotes the \temph{initial configuration}.
A \emph{transition} is an ordered pair of configurations, written $c \rightarrow c' \in C^2$.  If $c_p \ne c'_p$ for some $p \in P$ and $c'_q = c_q$ for every $q \ne p \in P$, the transition is a \temph{$p$-transition}.
A partial order  $\prec$ on local states induces a partial order on configurations, defined by $c \preceq c'$ if $c_p \preceq c'_p$ for every $p \in P$.  
\end{definition}
Normally, configurations over $P$ will have local states $X=S(P)$, but to define the notion of a grassroots system we will also use $X=S(P')$ for some $P' \supset P$.

\begin{definition}[Distributed Multiagent Transition System; Computation; Run]\label{definition:DMTS}
Given a set $X \subseteq S(\Pi)$ that includes $s0$ and a finite set of agents $P\subseteq \Pi$, a \temph{distributed multiagent transition system} over $P$ and $X$,
 $TS =(P,X,T,\lambda)$, has \temph{configurations} $C=X^P$; \temph{initial configuration}  $c0:=\{s0\}^P$;  a set of \temph{correct transitions} $T = \bigcup_{p \in P} T_p \subseteq C^2$, where each $T_p$ is a set of \temph{correct $p$-transitions}; and a \temph{liveness requirement} $\lambda \subset 2^T$ being a collection of sets of transitions.
A \temph{computation} of $TS$ is a potentially-infinite sequence of configurations over $P$ and $S$,  $r= c \xrightarrow{} c' \xrightarrow{}  \cdots $, with two consecutive configurations in $r$ referred to as a \temph{transition of $r$}.   A \temph{run} of $TS$ is a computation that starts with $c0$. 
\end{definition}

Multiagent transition systems are designed to address fault resilience, and hence need to specify faults, which are of two types -- safety (incorrect transitions) and liveness (infinite runs that violate a liveness requirement).  Def. \ref{definition:DMTS} summarizes what multiagent transition systems are, and Def. \ref{definition:ts-slc} below specifies safe, live and correct (safe+live) runs.  It also specifies the notion of a correct agent relative to a run.  All that an incorrect agent may do is violate safety or liveness (or both) during a run. 
Note that computations and runs may include  incorrect transitions.  The requirement that agent $p$ is live is captured by including $T_p$ in $\lambda$, and that all agents are live by  $\lambda = \{T_p : p \in P\}$.

\begin{definition}[Safe, Live and Correct Run]\label{definition:ts-slc}
Given a transition system  $TS=(P,X,T,\lambda)$, a computation $r$ of $TS$ is \temph{safe}, also $r \subseteq T$, if every transition of $r$ is correct.  We use  $c \xrightarrow{*} c' \subseteq T$ to denote the existence of a safe computation (empty if $c=c'$) from $c$ to $c'$. 
A transition $c'\rightarrow c''$ is \temph{enabled on $c$} if $c=c'$.
A run is \temph{live wrt $L \in \lambda$} if either $r$ has a nonempty suffix in which no transition in $L$ is enabled in a configuration in that suffix, or every suffix of $r$ includes a transition in $L$. A run $r$ is \temph{live} if it is live wrt every $L \in \lambda$.
A run $r$ is \temph{correct} if it is safe and live.
An agent $p \in P$ is \temph{safe in $r$} if $r$ includes only correct $p$-transitions;  is \temph{live in $r$} if for every $L \in \lambda$, s.t. $L \subseteq T_p$, $r$ is live wrt $L$; and  is \temph{correct} in $r$ if $p$ is safe and live in $r$. 
\end{definition}

Note that run is live if for every $L \in \lambda$ that is enabled infinitely often, the run has infinitely many $L$-transitions.  Also note that a finite run is live if for no $L\in \lambda$ there is an $L$-transition enabled on its last state.

A transition system is \emph{asynchronous} if progress by other agents cannot prevent an agent from taking an already-enabled transition.

\begin{definition}[Monotonic and Asynchronous Distributed Multiagent Transition System]\label{definition:multiagent-sa}
A distributed multiagent transition system $TS=(P,X,T,\lambda)$ is 
\temph{monotonic} wrt a partial order $\prec$ if $c\rightarrow c' \in T$ 
implies that $c \prec c'$, and it is \temph{monotonic} if it is monotonic wrt $\prec_S$.  It is \temph{asynchronous} if it is monotonic and for every $p$-transition $c \xrightarrow{} c' \in T_p$, $T_p$ also includes every $p$-transition $d \xrightarrow{} d'$ for which  $c \prec_{S} d$ and
    $(c_p \rightarrow c'_p) = (d_p \rightarrow d'_p)$.
\end{definition}

\subsection{Grassroots Protocols}  
Next, we define the notion of a protocol and a grassroots protocol.
\begin{definition}[Protocol]\label{definition:family}
A \temph{protocol} $\calF$ is a family of distributed multiagent transition systems over a local states function $S$ that has exactly one such transition system $TS(P) = (P,S(P),T(P),\lambda(P)) \in \calF$ over $P$ and $S$ for every
 $\emptyset \subset P \subseteq \Pi$.
\end{definition}

Note that in multiagent transition systems, safety and liveness are properties of runs, not of protocols.  Hence, when a protocol is specified with a multiagent transition system, there is no need to specify additional ``protocol specific'' safety or liveness requirements, as any such property is implied by the definition itself.

For simplicity and to avoid notational clutter, we often assume a given set of agents $P$, refer to the  member of $\calF$ over $P$ as a representative of the family $\calF$, and refer to the protocol $TS(P) = (P,S(P),T(P),\lambda(P)) \in \calF$ simply as $TS = (P,S,T,\lambda)$.

\begin{definition}[Projection]\label{definition:projection}
Let $\calF$ be a protocol over $S$, $\emptyset \subset P \subset P' \subseteq \Pi$.  Given a configuration $c'$  over $P'$ and $S(P)$, the \temph{projection of  $c'$ on} $P$, $c'/P$, is the configuration $c$ over $P$ and $S(P)$ satisfying $c_p = c'_p$ for all $p \in P$. The \temph{projection of $TS(P') = (P',S(P'),T',\lambda') \in \calF$ on $P$}, denoted $TS(P')/P$ is the transition system over $P$ and $S(P')$, $TS(P')/P:=(P,S(P'),T'/P,\lambda'/P)$,
where  $c_1/P \rightarrow c_2/P \in T'/P$ if  $c_1 \rightarrow c_2 \in T'$ and with $\lambda'/P := \{L/P : L \in \lambda'\}$.  
\end{definition}
Note that $TS(P')/P$ has local states  $S(P')$, not $S(P)$.  This is necessary as, for example, if the local state is a set of blocks, and in a $TS(P')$ configuration $c$ the local states of members of $P$ have blocks produced by members of $P'\setminus P$, this remains so also in $c/P$.

\begin{definition}[Subset]
Given  transition systems $TS=(P,X,T,\lambda)$,  $TS'=(P,X',T',\lambda')$, then $TS$ is a \temph{subset} of $TS'$, $TS \subseteq TS'$, if  $X\subseteq X'$, $T \subseteq T'$, and $\lambda$ is $\lambda'$ restricted to $T$,  $\lambda := \{ L \cap T |  L\in \lambda'\}$.
\end{definition}

\begin{definition}[Grassroots]\label{definition:grassroots}
A  protocol $\calF$ is \temph{grassroots} if $\emptyset \subset P \subset P' \subseteq \Pi \text{ implies that } TS(P) \subset TS(P')/P$. 
\end{definition}

Namely, in a grassroots protocol, a group of agents $P$, if embedded within a larger group $P'$, can still behave as if it is on its own (hence the subset relation), but also has new behaviors at its disposal (hence the subset relation is strict), presumably due to interactions between members of $P$ and members of $P'\setminus P$.

Here we define certain properties of a protocol used by Theorem \ref{theorem:grassroots}.
\begin{definition}[Monotonic and Asynchronous Protocol]\label{definition:monotonic-protocol}
Let $\calF$ be a protocol over $S$.
Then  $\prec_S$ is \temph{preserved under projection} if for every $\emptyset \subset P \subset P' \subseteq \Pi$ and every two configurations $c_1, c_2$ over $P'$ and $S$, 
$c_1 \preceq_S c_2$ implies that $c_1/P \preceq_S c_2/P$.
The protocol $\calF$ is \temph{monotonic} if $\prec_S$ is preserved under projection and every member of $\calF$ is monotonic; it is \temph{asynchronous} if, in addition, every member of $\calF$ is asynchronous.
\end{definition}

A protocol is interactive if a set of agents embedded in a larger group of agents  can perform  computations they cannot perform in isolation.  It is non-interfering if (\ia) \emph{safety:} a transition that can be carried out by a group of agents can still be carried out if there are additional agents that are in their initial state, and (\ib) \emph{liveness:} if an agent is live in a run of the group, then adding more agents to the group and extending the run with their transitions, will not result in the agent violating liveness. Formally:

\begin{definition}[Interactive \& Non-Interfering Protocol]\label{definition:non-interfering}
A protocol  $\calF$ over $S$ is \temph{interactive} if
for every $\emptyset \subset P \subset P' \subseteq \Pi$, 
$TS(P')/P \not\subseteq TS(P)$.
It is \temph{non-interfering} if for every $\emptyset \subset P \subset P' \subseteq \Pi$, with transition systems $TS = (P,S,T,\lambda), TS' = (P',S,T',\lambda') \in \calF$:
\begin{enumerate}
    \item \textbf{Safety}: For  every transition $c1 \rightarrow c2 \in T$,
$T'$ includes the transition  $c1' \rightarrow c2'$ for which $c1=c1'/P$, $c2=c2'/P$, and
$c1'_p = c2'_p = c0'_p$ for every $p \in P' \setminus P$, and
    \item \textbf{Liveness}: For every agent $p \in P$ and run $r$ of $TS$, if $p$ is live in $r$ then it is also live in every run $r'$ of $TS'$ for which $r'/P = r$.
\end{enumerate}
\end{definition}

\end{document}